\documentclass[aip,rsi,amsmath,amssymb,reprint]{revtex4-1}

\usepackage{graphicx}
\usepackage{dcolumn}
\usepackage{bm}
\def\ph2{{\it p}-H$_2$}

\begin{document}


\title{Computer simulation study of nanoscale size parahydrogen clusters}

\author{Massimo Boninsegni}
\email{m.boninsegni@ualberta.ca}
\affiliation{Department of Physics, University of Alberta, Edmonton, Alberta, T6G 2E1, Canada}

\date{\today}

\begin{abstract}
We present results of computer simulations of a parahydrogen cluster of a  thousand molecules, corresponding to approximately 4 nm in diameter, at temperatures between 1 K and 10 K. Examination of structural properties suggests that the local environment experienced by molecules is very similar to that in solid bulk parahydrogen, especially near the center of the cluster, where crystallization originates. 
Albeit strongly suppressed compared to helium, quantum-mechanical exchanges are not entirely negligible at the lowest temperature, resulting in a small but significant molecular mobility enhancement with respect to the bulk crystalline phase. 
Although the overall  superfluid response at the lowest temperature is only few percents, there is evidence of a surprising ``supersolid" core, as well as of a superfluid outer shell.  Much like in fluid parahydrogen at the melting temperature, quantum-mechanical signatures can be detected in the momentum distribution.

\end{abstract}

\maketitle


\section{\label{intro}Introduction}
Over four decades since the original prediction \cite{gs} of a possible superfluid transition  of a fluid of parahydrogen (\ph2) molecules at a temperature $T \approx 6$ K, its experimental observation continues to elude even the ablest practitioners and/or cleverest approaches. The main hurdle is, of course, the strong propensity of \ph2 to crystallize at temperatures significantly above those at which BEC and superfluidity (SF) ought to take place. Crystallization (hence {\em absence}\cite{note} of SF) is observed and/or predicted for bulk \ph2 even in physical settings in which an enhancement of the fluid phase might be expected, e.g., in reduced dimensions, disorder or confinement. \cite{bretz,schindler,sokol,boninsegni04,boninsegni05,screw,turnbull,boninsegni13,omiyinka,boninsegni16,delmaestro}\\ \indent 
This is, of course, a direct consequence of the interaction among two \ph2 molecules, significantly stronger than that, for example, between two atoms of helium, the archetypal superfluid. Indeed, recent theoretical calculations based on state-of-the art numerical techniques and realistic microscopic potentials suggest that, even if a metastable fluid phase of \ph2 could be stabilized considerably below freezing, it may not turn superfluid down to temperatures as much as two orders of magnitude lower than that of the original 1972 work. \cite{boninsegni18} 
\\ \indent 
On the other hand, quantitative theoretical predictions of superfluid behavior have been made for small clusters \cite{notesup} of \ph2 (few tens of molecules), both pristine \cite{sindzingre,fabio,fabio2,fabio3} as well as doped with molecules such as OCS \cite{kwon} or CO$_2$, \cite{castroni} at temperatures of the order of 1 K for pristine clusters (few tenths of a K for doped clusters). In some cases, the superfluid response of a \ph2 clusters has been predicted to be {\em enhanced} if clusters are confined in nanoscale size cavities with weakly adsorbing substrates.\cite{omi2}
For some of these predictions, claims of experimental verification have been made \cite{castroni,toennies}.  
\\ \indent 
As the number of \ph2 molecules increases, crystalline order begins to emerge, and concurrently the superfluid response of the cluster weakens, as molecules progressively enjoy lesser mobility than in smaller clusters.\cite{sup} Making allowance for the intrinsic ambiguity and imprecision of any such definition, given that one is dealing with systems of finite size, nevertheless it is meaningful to ascribe to clusters the qualities of ``solidlike'' or ``liquidlike'', based on the relative mobility of the elementary constituents.\cite{berry,charu} 
\\ \indent 
Virtually {\em all} microscopic calculations based on realistic intermolecular potentials suggest that \ph2 clusters comprising more than $N\sim 50$ molecules are solidlike, with no measurable superfluid response at $T$=1 K. \cite{fabio,fabio2,fabio3,khair,cuervo,guardiola} However, to our knowledge no systematic theoretical investigation has been carried out, aimed at assessing the emergence of crystalline order as the size of the cluster is increased. Some evidence of non-monotonic physical behavior, with possible re-entrance of liquidlike features, has been proposed in some studies, for some narrow ranges of sizes. \cite{cuervo}
More importantly, among the most significant recent experimental findings on \ph2 clusters there is that\cite{kuya} of Kuyanov-Prozument and Vilesov in 2008, who reported evidence of liquidlike behavior at a temperature $T$=1 K in \ph2 clusters of as many as 10,000 molecules, e.g., over two orders of magnitude greater than the theoretically expected maximum size.
\\ \indent 
The significance of such a result is twofold; on the one hand, it points to a highly nontrivial evolution of the physical properties of \ph2 clusters as a function of size, with possibly wide, reentrant fluid-like “phases” along the way to the equilibrium crystal. Secondly, it poses the question of a possible superfluid response of such a cluster, either in the form of ``bulk'' superfluidity, uniformly extended throughout the system as in smaller droplets, \cite{fabio4} or perhaps of a (few) superfluid  layer(s) on the surface of an otherwise  solid cluster. The latter scenario was explored by simulation in the context of a free \ph2 surface, but no conclusive evidence of superfluidity was obtained; \cite{wagner} it is conceivable that the curved geometry \cite{hernandez} of the cluster may alter some of the properties  of the outer layer, possibly enhancing molecular mobility. 
\\ \indent 
In this paper, we present results of a first principles computer simulation study of a \ph2 cluster comprising one thousand molecules, at temperature between 1 and 10  K. The purpose of this study is that of assessing the physical character of the system, in the light of the experimental findings of Ref. \onlinecite{kuya}. Particular attention is given to structural properties, to the occurrence of quantum-mechanical exchanges and, consequently, of superfluidity in the cluster. Our results show that a cluster of this size is solidlike, with molecules occupying preferential lattice positions and enjoying rather limited mobility. Though they are not suppressed to the same degree as in the crystalline phase of \ph2, nonetheless exchanges of molecules are infrequent, and as a result only a weak global superfluid response is predicted. 
\\ \indent 
There are nonetheless marked quantitative and qualitative differences between this system and bulk \ph2, chiefly a greater average mobility of molecules, not only near the surface of the cluster, as well as some distinct signatures of quantum-mechanical behavior associated with Bose statistics, detectable in the momentum distribution. While the overall superfluid response of the cluster is only few percents at the lowest temperature considered here, there are interesting indications of superfluidity at the surface as well as, somewhat unexpectedly, at the center of the cluster. The question of course arises as to which one of these two antithetical characters, if any, prevails in the ground state of the system, or whether perhaps this may be regarded as a ``nanoscale supersolid''. This issue can only be addressed by simulations carried out at considerably lower temperatures, which exceed the computational resources available for this project.
\\ \indent 
The remainder of this paper is organized as follows: in Sec. \ref{mm} the model of the system is introduced, and the computational methodology briefly reviewed; the results are presented in detail in Sec. \ref{res}; conclusions are outlined in Sec. \ref{conc}.
\section{Model and Methodology}\label{mm}
The system is described as an ensemble of $N=1,000$ point-like, identical particles with  mass $m$ equal to that of a \ph2 molecule, and with spin zero, thus  obeying Bose statistics.  The system is enclosed in a cubic cell, with periodic boundary conditions in the three directions; however, the size of the cell is taken sufficiently large (200 \AA), so as to make the boundary conditions irrelevant. Just like for smaller clusters, no special device is required (e.g., an {\em ad hoc} external potential) in order to ensure that the cluster stay together, this simply happens as a result of the intermolecular potential. 
\\ \indent 
The quantum-mechanical many-body Hamiltonian reads as follows:
\begin{eqnarray}\label{u}
\hat H = - \lambda \sum_{i}\nabla^2_{i}+\sum_{i<j}v(r_{ij})
\end{eqnarray}
where the first (second) sum runs over all particles (pairs of particles), $\lambda\equiv\hbar^2/2m=12.031$ K\AA$^{2}$, $r_{ij}\equiv |{\bf r}_i-{\bf r}_j|$ and $v(r)$ is a pair potential which describes the interaction between two molecules. We make use in this study of the accepted Silvera-Goldman (SG) pair potential, \cite{SG} which has been adopted in the vast majority of simulation studies of the condensed phase of \ph2, as well as \ph2 clusters.\cite{omibon}
\\ \indent 
We performed first principles QMC simulations of  the system described by Eq.  (\ref{u}), based on the continuous-space Worm Algorithm (WA), \cite{worm,worm2} based on Feynman's space-time approach to quantum statistical mechanics.\cite{spacetime}  Since this technique is by now fairly well-established, and extensively described in the literature, we shall not review it here. A canonical variant of the algorithm was utilized, in which the total number of particles $N$ is held fixed. \cite{fabio,fabio2}
Details of the simulation are  standard; for instance,  the short-time approximation to the imaginary-time propagator used here is accurate to fourth order in the time step $\tau$ (see, for instance, Ref. \onlinecite{jltp}). We have carried out numerical extrapolation of the estimates to the $\tau\to 0$ limit, and observed convergence for a value of $\tau=(1/320)$ K$^{-1}$.  All of the results quoted here were obtained using this value of time step. 
Simulations are started from a disordered initial arrangement of molecules resulting from a run at the highest temperature  considered here (10 K).
\\ \indent
Physical quantities of interest include, besides the energetics of the cluster at low temperature, structural correlations such as the radial density profile computed with respect to the center of mass of the system, which provide a direct indication of the propensity of the system to form an ordered, crystalline arrangement of molecules. Visual inspection of the many-particle configurations arising in the course of the simulation also offers valuable, qualitative insight. Finally, superfluid properties are investigated by computing the global \cite{sin1} and local \cite{paesani} superfluid density, as well as the one-body density matrix, from which the momentum distribution is obtained.
\section{Results}\label{res}
We begin by discussing the energetics of the cluster, featuring a very weak dependence on the temperature, in the range explored here (namely  $1$ K $\le T \le  10$ K), much as observed in the equilibrium crystal phase\cite{marisa} of \ph2. For example, the kinetic energy per molecule is 54.2(1) K at $T$=1 K, virtually unchanged, within the statistical uncertainties of the calculation, as the temperature is raised all the way to $T$=10 K. 
The energy per molecule at $T$=1 K is $-64.5(1)$ K, which can be compared against that ($\sim -88$ K) for bulk {\em hcp} \ph2 at the equilibrium density of 0.0261 \AA$^{-3}$ at the same temperature (the corresponding kinetic energy per molecule is $\sim 70$ K).\cite{omi3}
\begin{figure}[h]
\centering
\includegraphics[width=3.3in]{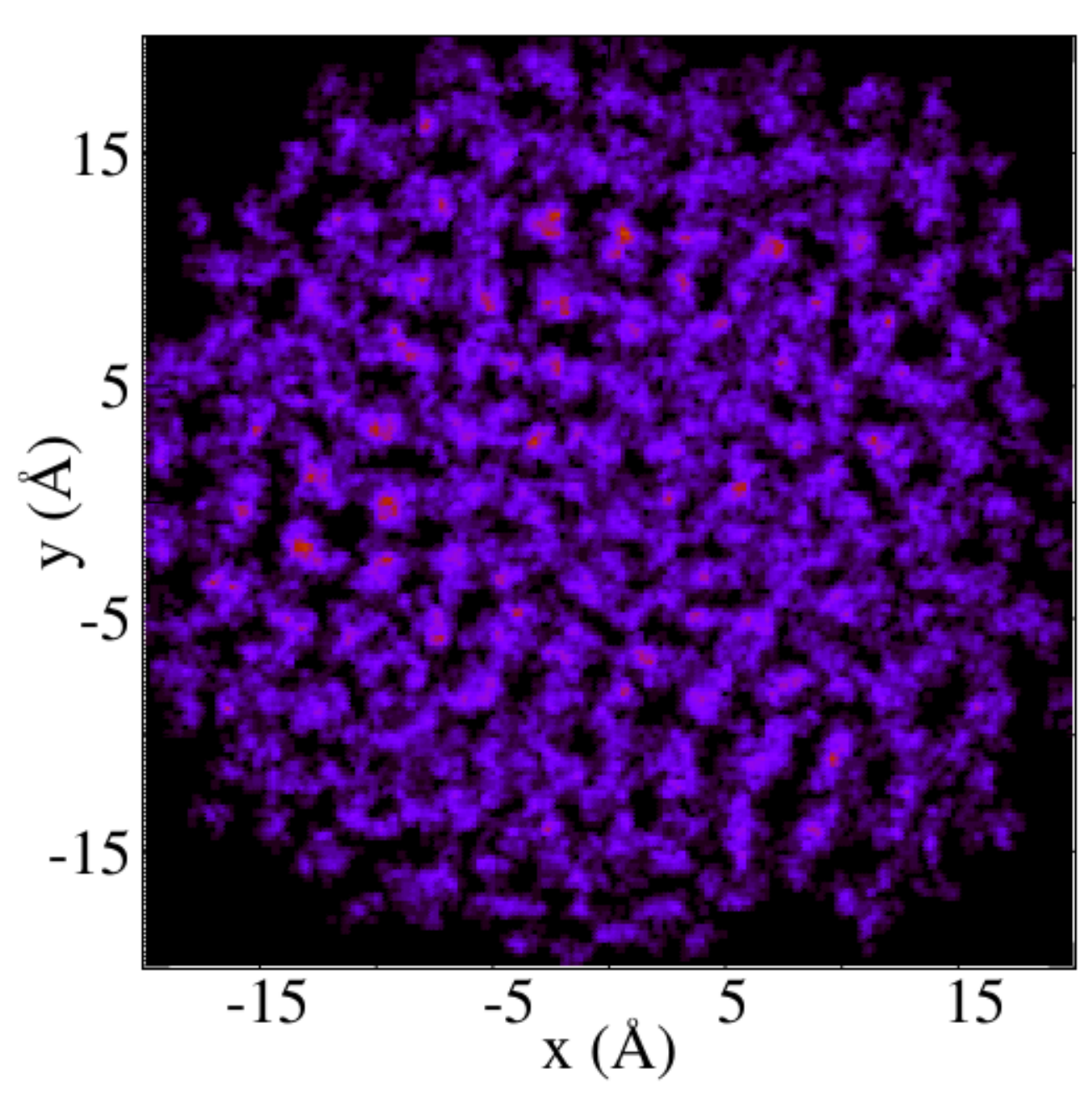}
\includegraphics[width=\linewidth]{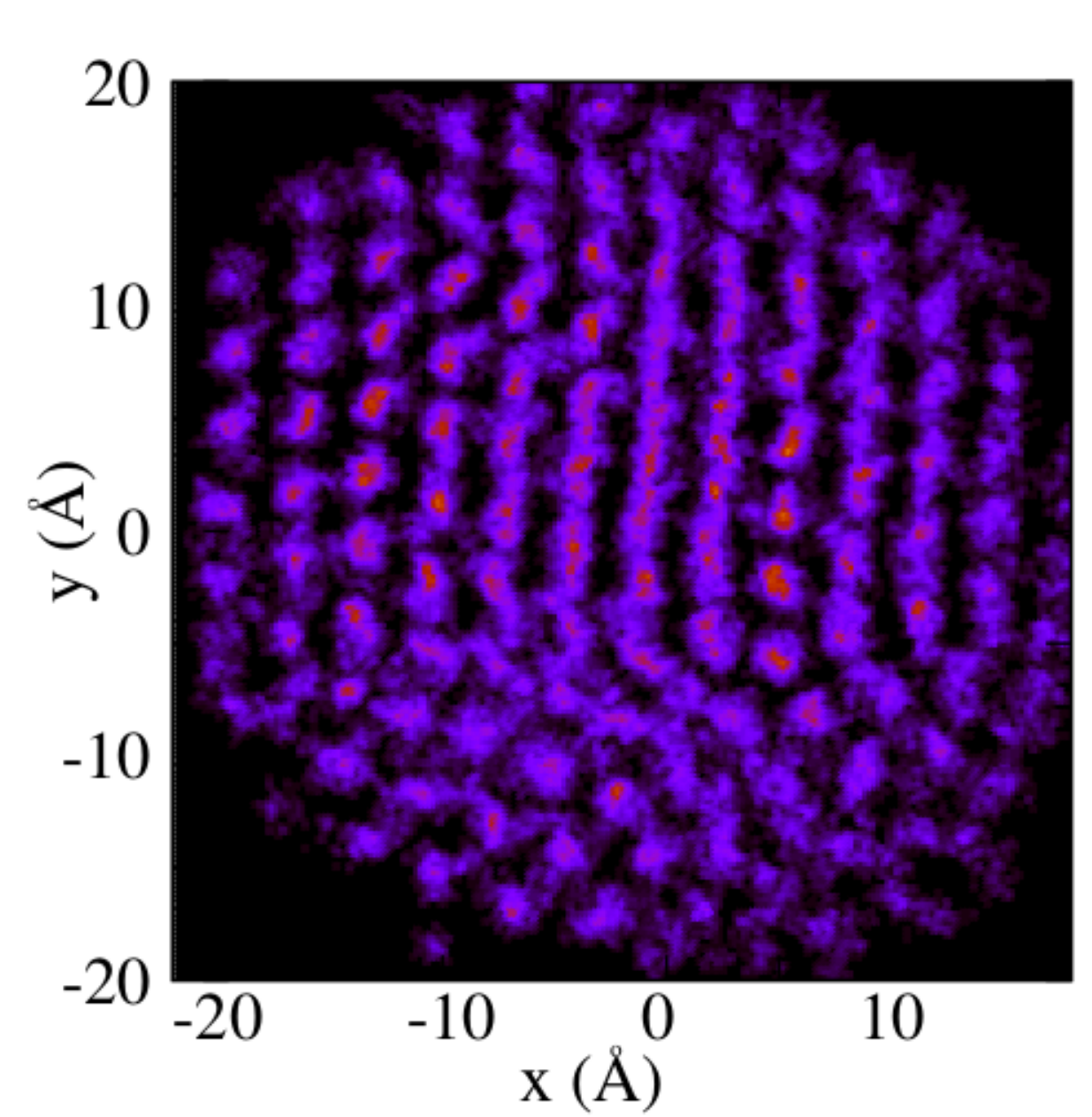}
\caption{{\em Color online.} Configurational snapshots of a cluster of $N=1000$ \ph2 molecules at $T$=8 K (upper panel) and $T$=4 K (lower panel). Shown in both cases is the \ph2 density integrated along the $z$ direction (brighter colors correspond to higher values). }
\label{f1}
\end{figure}
The significant difference in energy per molecule suggests that this cluster is still quantitatively far from bulk, and in particular that \ph2 molecules experience significantly less localization than in the crystalline bulk phase. This by itself is consistent with liquidlike behavior, but of course greater insight is afforded by a direct examination of the structure of the cluster.
\\ \indent 
Fig. \ref{f1} shows configurational snapshots for two runs at temperatures $T=8$ K (upper panel) and $T=4$ K (lower panel); what is shown is the \ph2 density integrated along one of the directions ($z$). The shape of the cluster is roughly spherical, the diameter being approximately 4 nm. The tendency of the molecules to arrange in an orderly fashion as the temperature is lowered from 8 to 4 K is clear. Obviously, these are merely configurational snapshots, i.e., on a different run the arrangement that molecules take spontaneously at the lower temperature will be different from that shown in the lower panel of Fig. \ref{f1}; however, they are offered here as representative of a physical trend, namely the strong tendency of the system to crystallize.
\begin{figure}[h]
\centering
\includegraphics[width=\linewidth]{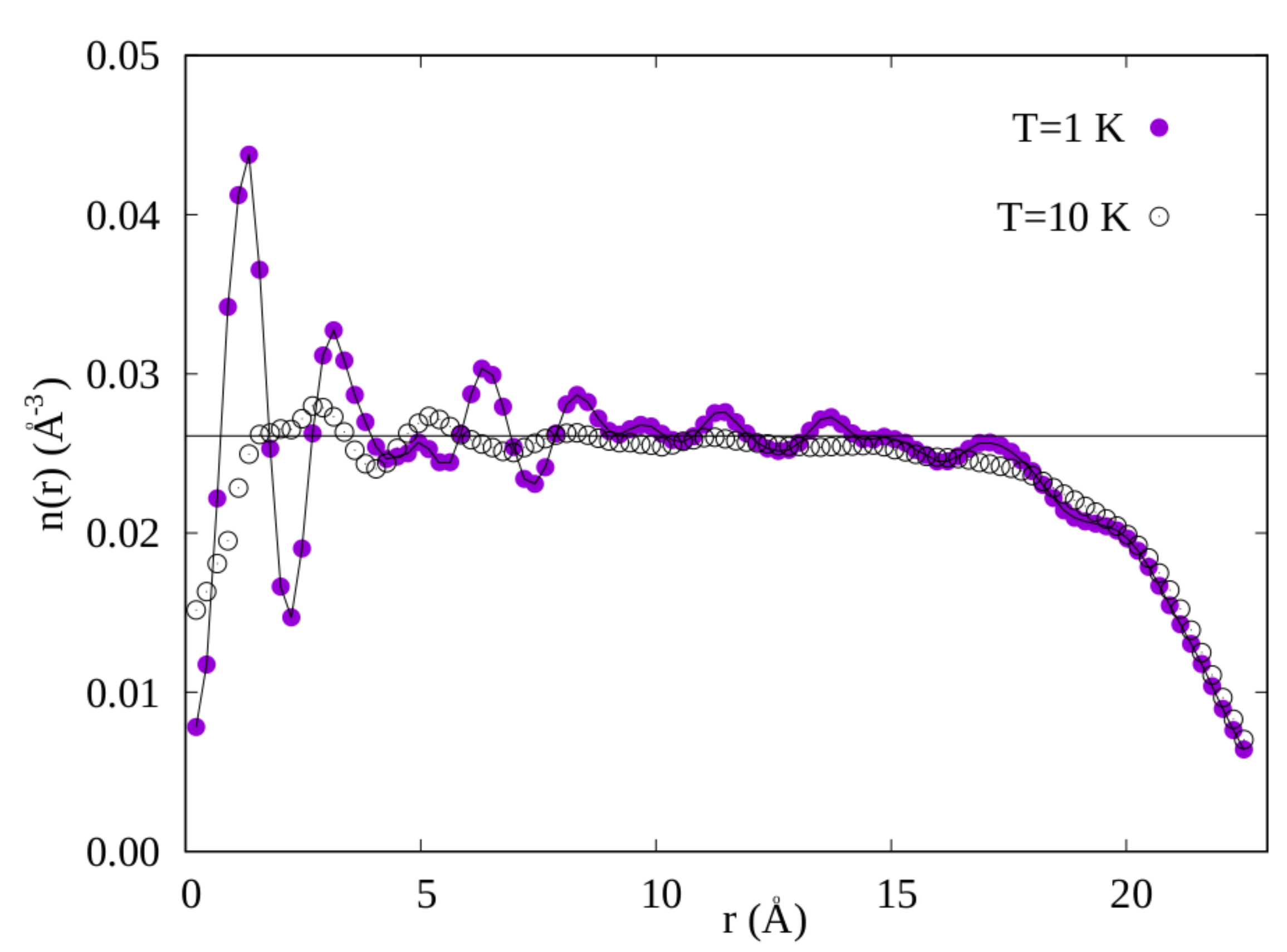}
\caption{{\em Color online.} Radial density profiles, computed with respect to the center of mass of the cluster, at temperature $T=1$ K (filled circles) and $T=10$ K (open circles). Statistical errors are smaller than symbol sizes. The line through the filled circles is a guide to the eye. The horizontal solid line corresponds to the equilibrium density of the {\em hcp} crystal, namely $n_\circ=0.0261$ \AA$^{-3}$. }
\label{f2}
\end{figure}
\\ \indent 
Fig. \ref{f2} shows radial density profiles (three-dimensional density $n(r)$ is expressed in \AA$^{-3}$) for the cluster at the two temperatures $T=1$ K (filled symbols) and $T=10$ K (open symbols). Also shown for reference (horizontal solid line) is the value of the bulk equilibrium density of the {\em hcp} crystal in the $T\to 0$ limit, namely $n_\circ=0.0261$ \AA$^{-3}$. The development of noticeable density oscillations at low temperature is consistent with the appearance of short-distance crystalline order, in line with the above, qualitative observation based on configurational snapshots. It is worth noting that the radial density oscillates around a value close to $n_\circ$, suggesting that a cluster of this size  rather closely reproduces, on a distance of few nm, the environment that \ph2 molecules experience in bulk crystal \ph2, especially near the center.  
\\ \indent 
At the same time, however, this system also displays significantly more marked quantum-mechanical effects than the bulk crystalline phase of \ph2 at the same temperature. The most direct measure of quantum delocalization of \ph2 molecules, is the average spatial extension of the world lines (paths in imaginary time),\cite{spacetime} defined as 
\begin{equation}
\delta = \sqrt{\langle[{\bf r}(\beta/2)-{\bf r}(0)]^2\rangle}
\end{equation}
where $\beta=1/k_BT$. Here, $\langle ...\rangle$ stands for thermal average, but averaging over all molecules is implied as well. At $T=1$ K, the value of $\delta$ averaged throughout the cluster is $\sim 1.5$ \AA, approximately 50\% greater than the corresponding value ($\sim 1$ \AA) in the equilibrium crystalline phase of \ph2.\cite{marisa} 
\\ \indent
Quantum-mechanical exchanges of molecules are rare in this system; however, they occur more frequently than in the bulk. For example, at a temperature $T=4$ K over 99.99\% of single-particle world lines close onto themselves, i.e., exchanges of identical molecules are virtually non-existent; as the temperature is lowered to $T=2$ K, still 99.97\% of world lines close onto themselves, and the rare permutation cycles that are observed involve at the most 6 molecules. On the other hand, at $T=1$ K 99.5\% of single-particle world lines close into themselves, and although permutations are still relatively rare, nonetheless cycles of exchanges comprising up to $\sim 30$ molecules are observed. So, a qualitative and quantitative change takes place at around $T=1$ K. It is worth noting that this frequency of exchanges, namely 0.5\%, albeit small is {\em orders of magnitude} greater than that observed in crystalline \ph2 at the same temperature.
\\ \indent
The global superfluid fraction of the cluster at $T=1$ K is approximately 2\%, an order of magnitude higher than that at $T=2$ K; the computed local superfluid density at $T=1$ K yields 
clear evidence of robust response of a
core shell ($ r \lesssim 5.5$ \AA), inside which the superfluid fraction averages\cite{expla}
 $\sim 50\%$. This intriguing observation suggests that such a shell, which comprises approximately 18 molecules, may display ``supersolid'' features,\cite{supersolid} given the above-mentioned clear presence of order at the center of the cluster. On the other hand, the bulk of the superfluid response of the cluster  appears to arise from a surface shell of thickness $\sim$ 2 \AA\  at a distance $r \gtrsim 20$ \AA\ from the center, consisting of approximately 200 molecules. The superfluid fraction in such an outer shell is about 7\%. This is consistent with the suggestion that a superfluid surface film may exist in a cluster of this size, although clearly the behavior at lower temperatures must be explored in order to make this conclusion firmer. The superfluid fraction in the rest of the cluster is significant (i.e, above statistical noise) only inside another shell of radius $\sim 2$ \AA, at a distance $r\sim 16$ \AA\ from the center.
\begin{figure}[h]
\centering
\includegraphics[width=\linewidth]{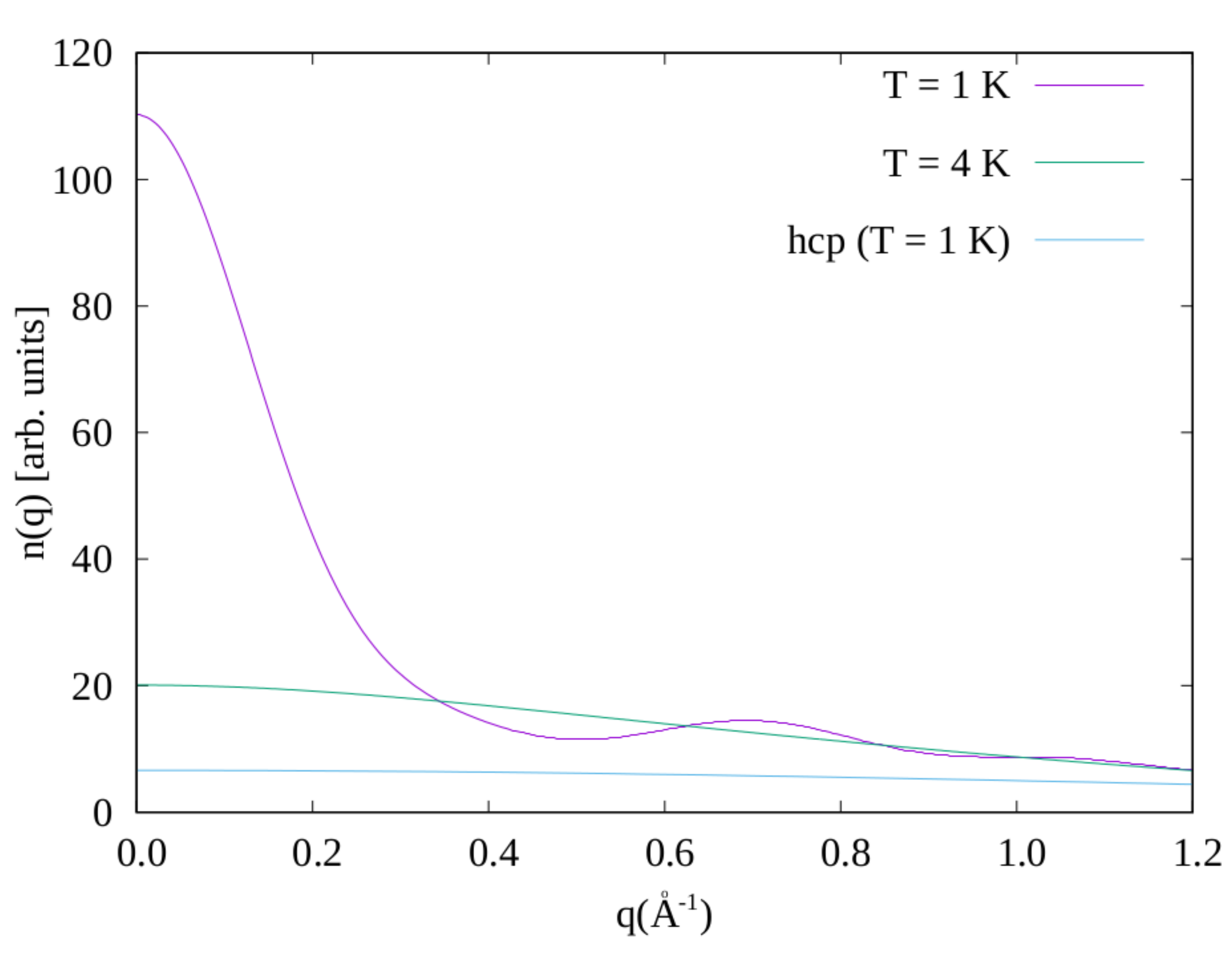}
\caption{{\em Color online.} Calculated momentum distribution of the droplet at temperature $T=1$ K and $T=4$ K. Statistical errors are too small to be shown on the scale of the figure. Also shown is the momentum distribution of the {\em hcp} crystal at $T=1$ K at the equilibrium density. }
\label{f3}
\end{figure} 
\\ \indent 
Additional insight into the dynamics of \ph2 molecules in the cluster is furnished by the momentum distribution $n(q)$, in principle experimentally accessible by neutron diffraction,\cite{sokol} although of course experiments on clusters are complicated by the low scattering cross section. Fig. \ref{f3} compares the  computed $n(q)$ for the cluster studied here at the two temperatures $T=1$ K and $T=4$ K. 
As the frequency of quantum-mechanical exchanges become non-negligible, at $T\sim 1$ K, the momentum distribution changes qualitatively, with the appearance of a peak at low $q$ which is characteristic of Bose-Einstein condensates. This is in stark contrast with the crystalline phase of \ph2, for which the momentum distribution (computed in a separate simulation for bulk \ph2) at $T=1$ K, shown in Fig. \ref {f3}, does not display any peak at low $q$ in the low temperature limit.
\section{conclusions}\label{conc}
Extensive, first principle quantum Monte Carlo simulations have been carried out of a \ph2 cluster of 1,000 molecules, in the range of temperature 1 K $\le T\le$ 10 K. The main goal of this study was gaining understanding on the structural and superfluid properties of a droplet of this size, following experimental observations suggesting that it might be liquidlike, and therefore possibly superfluid.\cite{kuya}
\\ \indent 
The conclusion of this study is that this cluster actually features both liquid and solidlike properties. Structurally, the emergence of crystalline order at low temperature, reminiscent of that in solid \ph2, is fairly clear; at the same time, however, quantum-mechanical exchanges of molecules are much more frequent than in solid \ph2, conferring to the system properties that may be regarded as liquidlike, at least insofar as they imply greater molecular mobility. Although the overall superfluid response of this cluster is rather small, nonetheless there is evidence of superfluidity at the surface, as well as, which is fairly surprising, at the core. 
\\ \indent 
The results obtained in this work do not rule out the possible scenario of 
a single superfluid layer on top of an otherwise insulating, crystalline drop, reminiscent of that of a $^4$He film adsorbed on a weak substrate, such as  Li.\cite{toigo} However, simulations at considerably lower temperature (easily an order of magnitude lower) will be necessary, in order to establish this conclusion more definitively, as the competition between the opposite tendencies to crystallize and superflow may result with the former prevailing in the low temperature limit.
In general, the results confirm that \ph2, even when it does escape (in part) crystallization, has considerable less propensity to turn superfluid than helium.\cite{boninsegni18} 
\\ \indent 
As a final remark, we discuss the melting of this cluster; it is known that thermal and quantum effects are roughly comparable in importance in the melting of bulk solid \ph2.\cite{marisa} Because quantum effects are found considerably more prominent than in the bulk, in a cluster of  this size, one expects the melting of this cluster to be largely quantum-mechanical in nature, and to occur at lower temperature than in the bulk (i.e., 13.8 K). Indeed, the results of this study suggest that the cluster is liquidlike, i.e., melted entirely, at $T\sim 8$ K.
\begin{acknowledgments}
This work was supported in part by the Natural Sciences and Engineering Research Council of Canada (NSERC). The author gratefully acknowledges the hospitality of the International Centre for Theoretical Physics, Trieste, where most of this research work was carried out. 
\end{acknowledgments}


\begin{thebibliography}{99}
\bibitem{gs}
V. L. Ginzburg and A. A. Sobyanin, JETP Lett. {\bf 15}, 242 (1972).
\bibitem{note}
There are neither theoretical predictions nor experimental evidence suggesting that a \ph2 crystal might display superfluid behavior. See, for instance,
A. C. Clark, X. Lin, and M. H. W. Chan
Phys. Rev. Lett. {\bf 97}, 245301 (2006).
\bibitem{bretz}
M. Bretz and A. L. Thomson, Phys. Rev. B {\bf 24}, 467 (1981).
\bibitem{schindler}
M. Schindler, A. Dertinger, Y. Kondo, and F. Pobell, Phys. Rev.
B {\bf 53}, 11451 (1996).
\bibitem{sokol}
P. E. Sokol, R. T. Azuah, M. R. Gibbs, and S. M. Bennington,
J. Low Temp. Phys. 103, 23 (1996).
\bibitem{boninsegni04}
M. Boninsegni, Phys. Rev. B {\bf 70}, 193411 (2004).
\bibitem{boninsegni13}
M. Boninsegni, Phys. Rev. Lett. {\bf 111}, 235303 (2013).
\bibitem{boninsegni05}
M. Boninsegni, New J. Physics {\bf 7}, 78 (2005).
\bibitem{turnbull}
J. Turnbull and M. Boninsegni,
Phys. Rev. B {\bf 78}, 144509 (2008).
\bibitem{boninsegni16}
M. Boninsegni, Phys. Rev. B 93, 054507 (2016).
\bibitem{screw}
M. Boninsegni, A. B. Kuklov, L. Pollet, N. V. Prokof'ev, B. V. Svistunov and M. Troyer, 
Phys. Rev. Lett. {\bf 99}, 035301 (2007). 
\bibitem{omiyinka}
T. Omiyinka and M. Boninsegni, Phys. Rev. B {\bf 93}, 104501 (2016).
\bibitem{delmaestro}
A. Del Maestro and M. Boninsegni, Phys. Rev. B {\bf 95}, 054517 (2017).
\bibitem{boninsegni18}
M. Boninsegni, Phys. Rev. B {\bf 97}, 054517 (2018).
\bibitem{notesup}
The superfluid fraction of a finite cluster is defined as $\rho_S=1-(I/I_{cl})$, where $I$ is the moment of inertia of the cluster with respect of the rotation axis, and $I_{cl}$ is its corresponding classical value. For details of its  evaluation in path integral Monte Carlo simulations, see Ref. \onlinecite{sin1}.
\bibitem{sin1}
P. Sindzingre, M. L. Klein and D. M. Ceperley, Phys. Rev. Lett. {\bf 63}, 1601 (1989).
\bibitem{sindzingre}
P. Sindzingre, D. M. Ceperley and M. L. Klein, Phys. Rev. Lett. {\bf 67}, 1871 (1991).
\bibitem{fabio}
F. Mezzacapo and M. Boninsegni, Phys. Rev. Lett. {\bf 97}, 045301 (2006).
\bibitem{fabio2}
F. Mezzacapo and M. Boninsegni, Phys. Rev. A {\bf 75}, 033201 (2007).
\bibitem{fabio3}
F. Mezzacapo and M. Boninsegni, J. Phys.: Condens. Matter {\bf 21}, 164205 (2009).
\bibitem{kwon}
Y. Kwon and K. B. Whaley
Phys. Rev. Lett. {\bf 89}, 273401 (2002).
\bibitem{castroni}
H. Li, R. J. Le Roy, P.-N. Roy and A. R. W. McKellar, Phys. Rev. Lett. {\bf 105}, 133401 (2010).
\bibitem{omi2}
T. Omiyinka and M. Boninsegni, Phys. Rev. B {\bf 90}, 064511 (2014).
\bibitem{toennies}
S. Grebenev, B. Sartakov, J. P. Toennies and A. F. Vilesov, Science {\bf 289}, 1532 (2000).
\bibitem{sup}
Specific clusters, e.g., (\ph2)$_{26}$ have been predicted to display simultaneously superfluidity and a geometrically regular ordering of molecules (Ref. \cite{supersolid}). 
\bibitem{supersolid}
F. Mezzacapo and M. Boninsegni, J. Phys. Chem. A {\bf 115}, 6831 (2011). 
\bibitem{berry}
R. S. Berry, T. L. Beck, and H. L. Davis, Adv. Chem. Phys. {\bf 70},75 (1988)
\bibitem{charu}
C. Chakravarty,
J. Chem. Phys. {\bf 103}, 10663 (1995). 
\bibitem{khair}
S. A. Khairallah, M. B. Sevryuk, D. M. Ceperley and J. P. Toennies, Phys. Rev. Lett.
{\bf 98}, 183401 (2007).
\bibitem{cuervo}
J. E. Cuervo and P.-N. Roy, J. Chm. Phys. {\bf 128}, 224509 (2008).
\bibitem{guardiola}
R. Guardiola and J. Navarro, Cent. Eur. J. Phys. {\bf 6}, 33 (2008).
\bibitem{kuya}
K. Kuyanov-Prozument and A. F. Vilesov, Phys. Rev. Lett. {\bf 101}, 205301 (2008).
\bibitem{fabio4}
F. Mezzacapo and M. Boninsegni, Phys. Rev. Lett. {\bf 100}, 145301 (2008).
\bibitem{wagner}
M. Wagner and D. M. Ceperley, J. Low Temp. Phys. {\bf 94}, 161 (1994).
\bibitem{hernandez}
E. S. Hernandez, M. W. Cole and M. Boninsegni, Phys. Rev. B {\bf 68}, 125418 (2003).
\bibitem{SG}
I. F. Silvera and V. V. Goldman, J. Chem. Phys. {\bf 69}, 4209 (1978).
\bibitem{omibon}
For recent refinements of the SG potential, see Ref. \onlinecite {omi3}.
\bibitem{omi3}
T. Omiyinka and M. Boninsegni, Phys. Rev. B {\bf 88}, 024112 (2013).
\bibitem{worm}
M. Boninsegni, N. Prokof'ev and B. Svistunov, Phys. Rev. Lett. {\bf 96}, 070601 (2006).
\bibitem{worm2}
M. Boninsegni, N. Prokof'ev and B. Svistunov, Phys. Rev. E {\bf 74}, 036701 (2006).
\bibitem{spacetime}
See, for instance, R. P. Feynman and A. R. Hibbs, {\em Quantum Mechanics and Path Integrals}, (McGraw-Hills, New York, 1965), Ch. 10.
\bibitem{jltp}
M. Boninsegni, J. Low Temp. Phys. {\bf 141}, 27 (2005).
\bibitem{paesani}
Y. Kwon, F. Paesani and K. B. Whaley, Phys. Rev. B {\bf 74}, 174522 (2006). 
\bibitem{marisa}
M. Dusseault and M. Boninsegni, Phys. Rev. B {\bf 95}, 104518 (2017).
\bibitem{expla}
It is difficult to furnish precise estimates of the local superfluid density in the $r\to 0$ limit, due to the limited statistics that can be accumulated in such a limited volume, in the course of a reasonable simulation.
\bibitem{toigo}
M. Boninsegni, M. W. Cole and F. Toigo, Phys. Rev. Lett. {\bf 83}, 2002 (1999).
\end{thebibliography}
\end{document}